\documentclass[conference]{IEEEtran}
\IEEEoverridecommandlockouts
\usepackage{cite}
\usepackage{amsmath,amssymb,amsfonts}
\usepackage{algorithmic}
\usepackage{graphicx}
\usepackage{textcomp}
\usepackage{xcolor}
\usepackage{multirow}
\def\BibTeX{{\rm B\kern-.05em{\sc i\kern-.025em b}\kern-.08em
    T\kern-.1667em\lower.7ex\hbox{E}\kern-.125emX}}
\begin{document}

\title{An FPGA Implementation of Displacement Vector Search for Intra Pattern Copy in JPEG XS\\

\author{\IEEEauthorblockN{Qiyue Chen, Yao Li, Jie Tao, Song Chen, Li Li, Dong Liu$^{\star}$}
University of Science and Technology of China, Hefei, China \\
\{qiyuechen, mrliyao, white\_tea0011\}@mail.ustc.edu.cn, \{songch, lil1, dongeliu\}@ustc.edu.cn}
\thanks{$\star$ Corresponding author: Dong Liu.}
}

\maketitle

\begin{abstract}

Recently, progress has been made on the Intra Pattern Copy (IPC) tool for JPEG XS, an image compression standard designed for low-latency and low-complexity coding. IPC performs wavelet-domain intra compensation predictions to reduce spatial redundancy in screen content. A key module of IPC is the displacement vector (DV) search, which aims to solve the optimal prediction reference offset. However, the DV search process is computationally intensive, posing challenges for practical hardware deployment. In this paper, we propose an efficient pipelined FPGA architecture design for the DV search module to promote the practical deployment of IPC. Optimized memory organization, which leverages the IPC computational characteristics and data inherent reuse patterns, is further introduced to enhance the performance. Experimental results show that our proposed architecture achieves a throughput of 38.3 Mpixels/s with a power consumption of 277 mW, demonstrating its feasibility for practical hardware implementation in IPC and other predictive coding tools, and providing a promising foundation for ASIC deployment.

\end{abstract}


\section{Introduction}
The Joint Photographic Experts Group (JPEG) committee recently developed the JPEG XS\cite{descampe2017jpeg,descampe2021jpeg} compression standard to meet the requirements of an international low-latency and low-complexity image compression standard. To enhance the performance of JPEG XS on remote desk\cite{herson2021remote,wharton2022optimizing} and keyboard-video-mouse (KVM) applications\cite{richter2023jpeg,rodrigues2024low,nakachi2025privacy}, several techniques have been developed recently to improve its coding efficiency on screen content\cite{xu2021overview,nguyen2021overview}, such as Temporal Differential Coding (TDC)\cite{bruylants2024tdc}, which performs temporal predictions between adjacent frames, and Intra Pattern Copy (IPC)\cite{li2025frequency}, which performs intra-copy predictions within a single frame. The IPC utilizes a wavelet-domain prediction framework to reduce the image structural redundancy, achieving notable BD-PSNR improvement on screen content.

In IPC, the coefficient compensation predictions are performed in a group-based manner, with each group corresponding to the coefficients of a spatial position in several decomposition bands.
Wavelet coefficients are grouped from both the spatial and frequency domains. Each coefficient group is predicted by the reference coefficients from the reconstructed image, according to a displacement vector (DV)\cite{nagel1983constraints,nagel1986investigation,liu2024vector}.
The IPC prediction loop has four main parts: DV search, pattern compensation, mode selection, and reconstructed precinct buffer. 

Among the above modules, DV search is the most computationally intensive, resource-intensive, and delay-sensitive, as it traverses all candidates to determine the optimal one. 
Due to the need for iterating all possible prediction offsets and solving the DV that can minimize the overall coding cost to retrieve the final prediction reference, this module becomes a critical bottleneck in terms of hardware complexity and latency, which hinders the practical deployment of IPC in real-time hardware systems\cite{yang2022fpga,tian2024implementation,alcain2021hardware}. Extensive FPGA and ASIC implementations have been developed for motion estimation and intra prediction in H.264 and HEVC, including array-based block-matching engines with SAD/SATD cost evaluation, hierarchical or TZ-search methods, and multi-block architectures optimized for throughput and bandwidth efficiency \cite{shajin2022efficient,ahmad2024efficient}. These designs exploit spatial or temporal redundancy between pixel blocks and rely on regular memory access and fixed block partitions.

However, none of these works operate on the low-latency and low-complexity JPEG XS framework or support the grouped, frequency-domain prediction flows specific to JPEG XS IPC. Therefore, we propose an optimized FPGA-based implementation of the DV search module, enabling efficient IPC hardware realization and establishing a foundation for future ASIC deployment. The main contributions of this paper are summarized as follows.

1) FPGA architecture design for DV search in the IPC framework. This paper, for the first time, proposes an FPGA architecture for the DV search module in the JPEG XS IPC framework. It introduces a four-stage pipelined design that balances throughput and latency, enabling parallel residual computation and DV comparison across multiple IPC groups.

2) Optimized memory organization for the DV search architecture. To address the scattered patterns of wavelet coefficients, we develop a novel memory organization that aligns wavelet coefficients by IPC Groups and Units, coupled with an on-chip Translation Lookaside Buffer (TLB) to store variable block sizes.

The rest of this paper is organized as follows. Section II introduces the frequency domain IPC framework and its DV search methods. Section III details the proposed DV search hardware architecture. Section IV shows the experimental results. Finally, Section V concludes this paper.

\begin{figure*}[htbp]
	\centerline{\includegraphics[width=\textwidth]{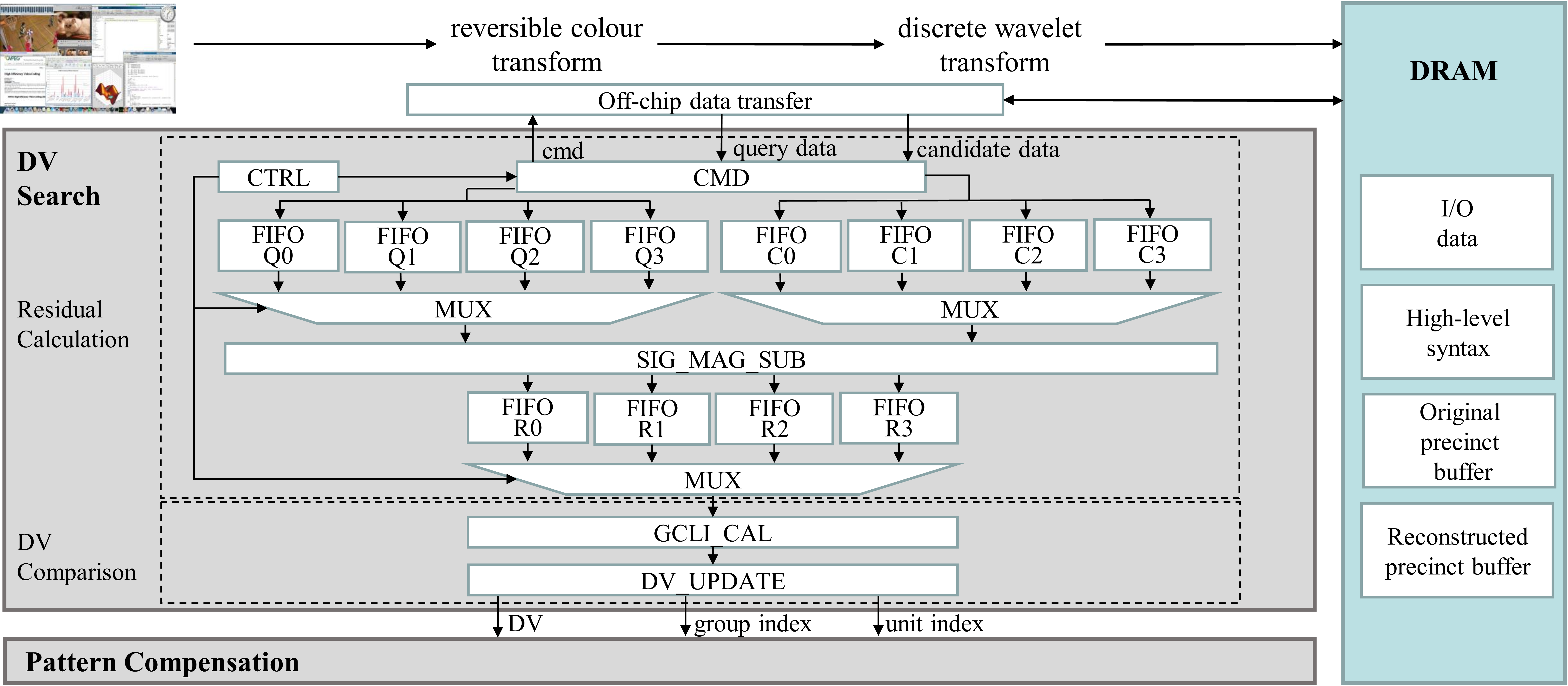}}
	\caption{Proposed DV search system architecture. The system is composed of residual calculation and the DV comparison engine. The residual calculation engine retrieves IPC Unit data from DRAM, computes residuals and forwards them to the DV comparison engine, which selects the optimal DV based on the estimated bit plane count of residuals and passes it to the subsequent pattern compensation module.}
	\label{fig:system_architecture}
\end{figure*}

\section{Background}

The IPC\cite{li2025frequency} is designed to improve the coding efficiency of screen content on JPEG XS. It performs coefficient group-based intra compensation predictions, with coefficients grouped into IPC Unit and IPC Group from spatial and frequency dimensions, respectively.
This two-level grouping scheme ensures that coefficients within the same group share similar frequency characteristics and thus share the same DV for synchronized compensation. 

For each IPC Group, the DV search module traverses candidate patterns and retrieves reference coefficients from the co-indexed bands of reconstructed precincts. Residuals are calculated by subtracting predictions from original values. The optimal DV is selected from available prediction candidates by comparing the bitplane count of the residuals. Finally, the encoder chooses between IPC and original coding based on the estimated overall coding cost.

The DV search module involves high computational complexity, which poses challenges for practical hardware implementation\cite{shi2022fpga,kong2022software,sidler2017accelerating}. Therefore, we propose a pipelined FPGA architecture design for DV search and optimize the memory organization to facilitate its deployment, which will be detailed in the following section.

\section{Proposed Pipelined DV Search Architecture}

In this section, we propose a pipelined hardware architecture to accelerate the DV search process in the IPC framework. First, we present the overall top-level architecture, followed by a detailed introduction to the two main processing engines: the residual calculation engine and the DV comparison engine. Second, we illustrate the external memory organization that supports efficient data access during the search process.

\subsection{Top Architecture}

The top-level structure for the pipelined DV search is illustrated in Fig. 1. The inputs are original and reconstructed wavelet coefficients from the JPEG XS codec, which are generated by the reversible color transform (RCT) and discrete wavelet transform (DWT) to facilitate its deployment. These coefficients are then partitioned and stored in two IPC unit banks on DRAM. The original IPC unit bank stores the original wavelet coefficients of the original precinct, while the reconstructed IPC unit bank stores the reconstructed wavelet coefficients of the reconstructed precinct.

The DV search architecture has two components: a residual calculation engine and a DV comparison engine. The residual calculation engine retrieves IPC Units from memory and computes block-wise residuals. The DV comparison engine evaluates the bit cost of each residual IPC Unit and searches for the optimal matching DV, thus completing the DV search loop. The key modules of the architecture are listed below.

1) CTRL: CTRL controls the CMD module by determining which first-in-first-out (FIFO) buffer the original and reconstructed data should be stored in. It also ensures that original and reconstructed data are read simultaneously from the two FIFOs based on the group index to be passed to SIG\_MAG\_SUB.

2) CMD and Off-chip data transfer: CMD is used to map the precinct number to the memory address. It also calculates the commands related to each memory address and receives original and reconstructed data from Off-chip data transfer. The off-chip data transfer module takes commands from CMD and accesses DRAM through the customized interface.

3) SIG\_MAG\_SUB: SIG\_MAG\_SUB reads original and reconstructed IPC Units from two FIFOs, computes signed residuals and outputs residuals. These residuals are then routed to residual FIFOs according to the group index.

4) GCLI\_CAL: This module calculates the residual bit cost. Based on the group index, CTRL routes residuals from four FIFOs to GCLI\_CAL via a shared MUX. For each group, GCLI\_CAL performs a bitwise OR on all residuals and computes the corresponding GCLI cost. After processing all groups, it accumulates the GCLI overhead to obtain the total bit cost, which is then forwarded to DV\_UPDATE.

5) DV\_UPDATE: DV\_UPDATE updates the optimal DV for the current grouping and unit position based on the calculated bit cost, to ensure the selection of the optimal DV combination in the DV comparison process. This process compares the current and historical minimum costs and updates the optimal solution when a smaller cost is achieved.

\subsection{Residual Calculation Engine}

The primary function of the residual calculation engine is to perform block-wise residual calculation. 
It retrieves a collection of original and reconstructed coefficient blocks from various bands via the CMD module. After retrieval, coefficients are streamed into on-chip query and candidate FIFO arrays: Q0–Q3 for original data and C0–C3 for reconstructed data, where 0-3 corresponds to 4 IPC Groups.
The CTRL module manages these FIFOS and ensures synchronized read/write operations. When a FIFO is underfilled, the CTRL triggers a full burst read from DRAM via the CMD module.

Once both query and candidate FIFOs are ready for an IPC Group, the MUX modules select a pair of coefficients and put them to the SIG\_MAG\_SUB module. Each coefficient is a 32-bit signed value in sign-magnitude format and is subsequently split into sign and magnitude fields. Four parallel subtraction paths are implemented to compute signed residuals for different sign combinations. The computed residuals are stored in residual FIFOs (R0–R3) and then routed to the GCLI\_CAL module through a shared MUX, maintaining group-wise data alignment.

\subsection{DV comparison Engine} 

Fig. 2 depicts the proposed four-stage DV comparison hardware architecture. The architecture accommodates DV comparison at the block level by systematically piping each processing step, which enables parallel DV evaluation with low latency and supports scalable block-wise processing. It consists of data processing modules (white blocks) and registers (grey blocks) that store intermediate results across pipeline stages.

\begin{figure}[htbp]
	\centerline{\includegraphics[width=\linewidth]{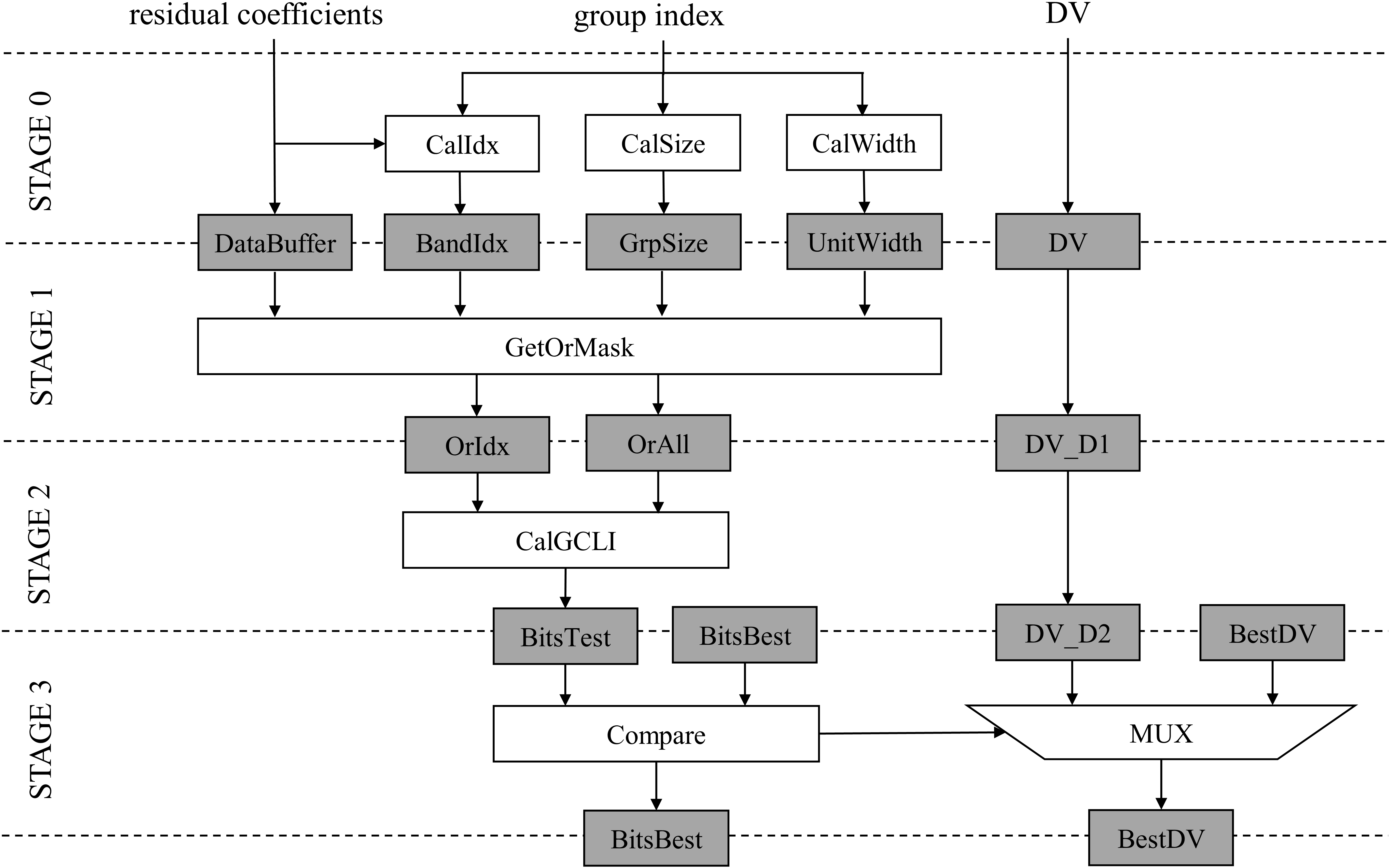}}
	\caption{Proposed four-stage DV comparison hardware architecture. The pipeline consists of four processing stages with data processing and register synchronization.}
	\label{fig: DV comparison}
\end{figure}

In stage 0, the architecture takes residual coefficients, group index, and DV as inputs. Residual coefficients are loaded into the data buffer, while calculation models(CalIdx, CalSize, CalWidth) generate parameters including Bandidx, Grpsize, and Unitwidth based on the current group index. These parameters configure the pipeline to support flexible group-wise calculation, adapting to block widths of different groups.

In stage 1, the GetOrMask module collects the current residual data and group data (BandIdx, GrpSize, UntWidth) to compute a bitwise OR mask across the group. The outputs OrIdx and OrAll are stored along with the delayed DV (DV\_D1) for stage 2.

In stage 2, the CalGCLI module performs bit-level GCLI calculation by referencing OrIdx and OrAll. It generates BitsTest for the current residual coefficient group. Simultaneously, DV\_D1 is further delayed to DV\_D2, aligning it with BitsTest for the comparison stage.

In stage 3, the Compare module compares the BitsTest result with the previously recorded BitsBest and updates BitsBest when a smaller BitsTest is achieved. A MUX is used to select the best DV based on the results of the Compare module. The final outputs, including BestBits and BestDV, are passed to the subsequent modules for pattern compensation. 

\subsection{External Memory Organization}

\begin{figure}[htbp]
	\centerline{\includegraphics[width=\linewidth]{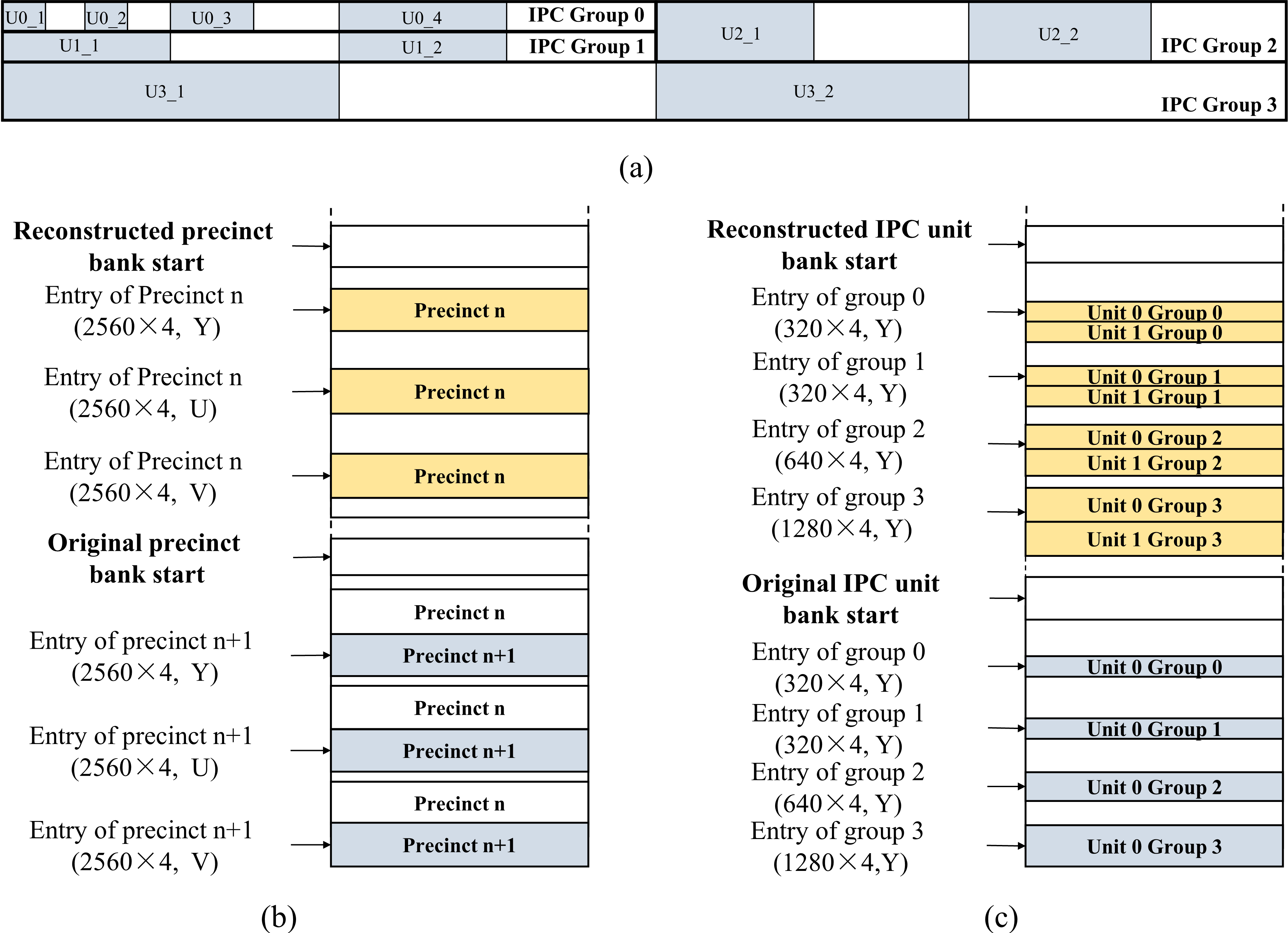}}
	\caption{(a) Relationship between the IPC Group and IPC Unit, illustrated with a five horizontal, two vertical decomposition. Blue blocks denote IPC unit 0 across different IPC Groups and sub-bands. The notation \(Ui\_j\) denotes the \(i\)th IPC Unit within the \(j\)th IPC Group. (b) Method 0: Precinct-aligned memory organization. Each precinct has a fixed size of $2560 \times 4$, where 2560 and 4 denote the precinct's width and height, respectively. (c) Method 1: IPC Group-aligned memory organization. The block sizes of IPC groups vary according to a 5-horizontal, 2-vertical decomposition.}
	\label{fig: memory mapping}
\end{figure}

\begin{figure}[htbp]
	\centerline{\includegraphics[width=\linewidth]{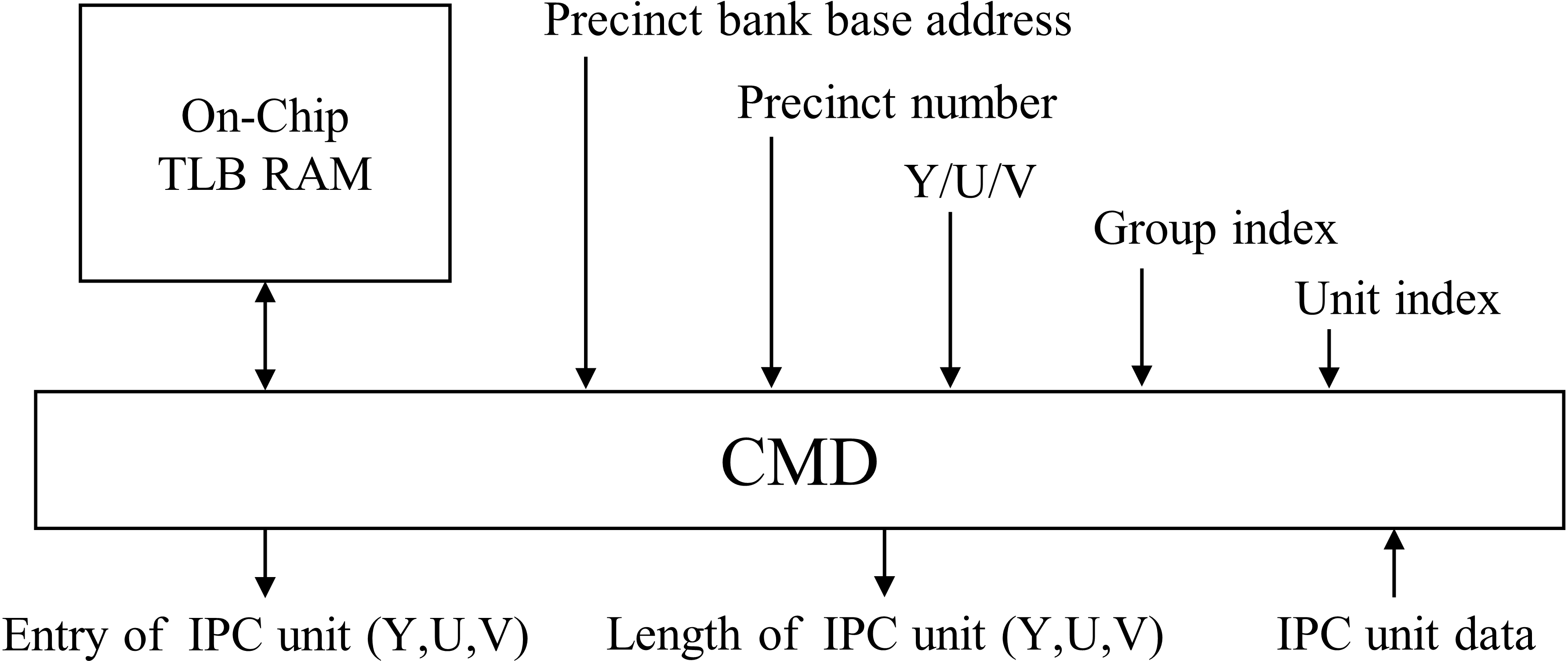}}
	\caption{External memory addressing. }
	\label{fig: memory addressing}
\end{figure}

Fig. 3(a) shows the relationship between IPC Groups and wavelet coefficient blocks from various bands within an IPC Unit. A naive approach of memory organization is depicted in Fig. 3(b), referred as Method 0. It stores reconstructed and original coefficients in two precinct-based memory banks, arranged sequentially in DRAM. 
Although offering a straightforward linear addressing mechanism, Method 0 is inefficient for IPC's group-based access because the scattered coefficient blocks from different sub-bands are scattered and must be individually located by group, unit, and band index, which will add control complexity and lower memory throughput.

To address these issues, we proposed a novel memory organization manner, referred as Method 1 (Fig. 3(c)), which organizes coefficients by IPC Groups and Units instead of precincts. IPC Units in the same group are stored sequentially, with each group entry containing all sub-band blocks. This allows loading all blocks of an IPC Unit using a single base address plus a fixed offset, eliminating the need for separate address computation and facilitating burst-based memory access.

The external memory addressing is realized by the CMD module, as shown in Fig. 4. 
An on-chip TLB RAM stores the length of each coefficient block within IPC Units across different groups, which differs according to IPC group index.
The CMD module sequentially traverses original and reconstructed wavelet coefficients based on the group and unit index, and reads the IPC Unit at each corresponding entry. 
Entry addresses are generated based on the precinct bank base address, precinct number, Y/U/V selector, group index, unit index and the on-chip TLB RAM, which is updated whenever the DV search process transitions to the next precinct.

\section{Experimental Results}

Implemented on a Xilinx Artix-7 (XC7A35T) at 100 MHz, our FPGA design achieves rate–distortion performance consistent with the IPC reference software while maintaining a low latency of 73.01 ms. The FPGA resource utilization of the proposed architecture is summarized in Table \ref{tab-1}. Among the modules, the DV comparison engine, composed of GCLI\_CAL and DV\_UPDATE, accounts for the majority of logic consumption. Specifically, the residual cost calculation module GCLI\_CAL dominates the area with 11.63 K LUTs, 19.98 K FFs, and 17 DSPs due to its intensive computation of coding costs across all candidates. The residual calculation engine occupies 0.48 K LUTs and 0.47 K FFs, requiring 15 BRAMs to buffer wavelet coefficients.

\begin{table}[t]
	\renewcommand\arraystretch{1.1}
	\centering
	\footnotesize
	\caption{Resource Utilization Results}
	\label{tab-1}
	\setlength{\tabcolsep}{1.6mm}
	\begin{tabular}{l|c|cc}
		\hline
		\multirow{2}{*}{Module} 
		& \multirow{2}{*}{Residual Calculation} 
		& \multicolumn{2}{c}{DV Comparison} \\
		&  & GCLI\_CAL & DV\_UPDATE \\
		\hline
		LUTs (K) & 0.48 & 11.63 & 0.73 \\
        FFs (K) & 0.47 & 19.98 & 1.41 \\
        DSPs & 0 & 17 & 0 \\
        BRAM & 15 & 0 & 0 \\
		\hline
	\end{tabular}
\end{table}

To evaluate the proposed memory organization, we compare the baseline Method 0 and the optimized Method 1 under identical conditions. As summarized in Table \ref{tab-2}, both designs run on the same FPGA platform at 100 MHz. Method 1 increases the throughput to 38.30 Mpixels/s while maintaining comparable power consumption, yielding a 6.1\% improvement in power efficiency (138.27 Mpixels/s/W). 
In terms of resources, the usage of LUT and FF is reduced by 7. 5\% and 8. 4\%, respectively, with a moderate increase in BRAM. The superior performance and efficient resource utilization demonstrates the effectiveness of the proposed memory-optimized architecture.

\begin{table}[t]
\renewcommand\arraystretch{1.15}
\centering
\footnotesize
\caption{Performance and Resource Comparison of Pattern Search}
\label{tab-2}
\setlength{\tabcolsep}{2mm}
\begin{tabular}{l|cc}
\hline
\textbf{Parameter} & \textbf{Method 0 (Baseline)} & \textbf{Method 1 (Proposed)} \\
\hline
\multicolumn{3}{l}{\textit{Performance Metrics:}} \\ 
\hline
Platform & \multicolumn{2}{c}{Xilinx Artix-7 (XC7A35T), 100 MHz} \\
Throughput (Mpixels/s) & 35.98 & 38.30 \\
Power (mW) & 276 & 277 \\
Power Efficiency (Mpixels/s/W) & 130.36 & 138.27 \\
\hline
\multicolumn{3}{l}{\textit{Resource Utilization:}} \\
\hline
LUTs (K) & 13.93 & 12.89 \\
FFs (K) & 23.80 & 21.79 \\
DSPs & 17 & 17 \\
BRAM & 11 & 15 \\
\hline
\end{tabular}
\end{table}

\section{Conclusion}

 This paper presents for the first time an efficient FPGA-based design of the DV search module for the IPC framework in JPEG XS, aiming to promote the practical hardware implementation of the computationally intensive DV search module. A four-stage pipelined architecture with optimized memory organization is proposed to ensure high-throughput residual calculation and DV comparison, with a four-stage DV comparison pipeline to enhance matching efficiency across coefficient groups. Experimental results show that our system achieves a throughput of 38.3 Mpixels/s with a power consumption of 277 mW, demonstrating its efficiency for hardware-oriented wavelet-domain compensation prediction and providing a practical reference for power-constrained and real-time IPC deployment on ASIC. 

\bibliographystyle{IEEEtran}
\bibliography{reference}

\end{document}